\documentclass[prl,aps,showpacs,twocolumn]{revtex4}
\usepackage{graphicx}      
\newcommand{\be}{\begin{eqnarray}}
\newcommand{\ee}{\end{eqnarray}}
\def\beq{\begin{equation}}
\def\eeq{\end{equation}}

\begin{document}
\title{Finite size corrections to the blackbody radiation laws}
\author{Antonio M. Garc\'{\i}a-Garc\'{\i}a}
\affiliation{Physics Department, Princeton University, Princeton, New Jersey 08544, USA}
\affiliation{The Abdus Salam International Centre for Theoretical
Physics, P.O.B. 586, 34100 Trieste, Italy}
\begin{abstract}
We investigate the radiation of a blackbody in a cavity of finite size.
For a given geometry,  we use semiclassical techniques to obtain explicit expressions of the modified Planck's and Stefan-Boltzmann's 
blackbody radiation laws as a function of the size and shape of the cavity.    
We determine the range of parameters (temperature, size and shape of the cavity) 
for which these effects are accessible to experimental verification.   
Finally we discuss potential applications of our findings in the physics of the cosmic microwave background  and sonoluminescence. 
\end{abstract}
\maketitle 
To a good approximation a blackbody can be represented by a cavity heated to a temperature $T$ and connected to the outside by a small
aperture.  
%The resulting
%blackbody spectrum may be detected by measuring the energy flux through the aperture.
A useful quantity to describe the radiation properties of a blackbody is
the density of energy per unit of frequency,
\be
\label{a1}
u(\nu) = \frac{\rho(\nu)}{V} \epsilon (\nu)
\ee 
where $V$ is the volume of the blackbody cavity, $\rho (\nu) d\nu$ is the number of stationary electromagnetic
modes with frequencies between $\nu$ and $\nu + d\nu$ and $\epsilon (\nu)$ is the average energy per mode.
In the $V \to \infty$ limit, 
\be
 \rho(\nu) = \rho_V(\nu) = 
{V}\frac{8 \pi \nu^2}{c^3}  \nonumber
\ee
where $c$ is the speed of light.
Classically the average energy 
per mode does not depend on the frequency $\nu$ so, $u(\nu) \propto \nu^2$. However  
experimental results indicate 
$u(\nu) \rightarrow 0$,  $\nu \rightarrow \infty$. 
Planck noted that agreement with experiments could be achieved
if the energy was considered a discrete variable,
\be 
 \epsilon (\nu) = h\nu /\left( e^{\frac{h\nu}{k_B T}}-1
\right) \nonumber
 \ee 
($k_B$ and $h$ are the Boltzmann and Planck constants
respectively). 
Combining these two results, Planck's radiation law  
%defined as the energy per unit volume of the cavity
 follows immediately,
%law
$
u _V(\nu) = \frac{\rho_V(\nu)}{V}\frac{h \nu}{e^{h \nu/k_B
T} - 1}.
$
Another useful quantity to characterize the blackbody radiation is the total energy
emitted per unit of time and area,
\be
 R(T) = {c \over{ 4\pi}}\int_0^\infty d\nu u(\nu). \nonumber
\ee
In the $V \to \infty$ limit it is given by the Stefan-Boltzmann law,  
\be 
R(T) = \sigma T^4 \nonumber
\ee
where $\sigma = 5.67 \times 10^{-8}  	J s^{-1} m^{-2}K^{-4}$.  
%This relation  thus describes the equilibrium properties of matter
%and radiation, and is 
The blackbody radiation, besides its role as a precursor of the quantum theory,
has been an invaluable tool in a variety of applications. For example,
it opened the way to the thermal remote sensing of bodies that allows to
diagnose the properties of materials or to measure the surface temperature of
celestial objects. The international scale of temperature, in the range above
the freezing point of Silver, is based on Eq.~(\ref{a1}).

Finite size corrections to these results are expected since 
$\rho_V(\nu)$  is only the leading term of a full expansion 
 in powers of a typical length $L$ of the cavity. 
A comprehensive theoretical study of $\rho(\nu)$ was carried out in \cite{bal,blo1} by 
 using the Green function formalism previously developed for scalar waves in Refs.\cite{blo,blo2}. 
 In the special case of 3d cavities with smooth surfaces and $L \gg c/\nu$,
\begin{equation}
\label{den2}
\rho(\nu) = \sum_i \delta(\nu-\nu_i) =  \rho_V (\nu)+  \rho_{\cal C} +{\tilde \rho}(\nu), 
\end{equation}
where $\nu_i$ stands for the $ith$ natural mode of the cavity,  
$\rho_{\cal C}= {2 \over 3\pi}{ {\cal C} \over c}$ and  ${{\cal C} \propto L}$ is the mean curvature (${\cal C} = 4\pi r_0$ for a sphere \cite{baduri} of radius $r_0$), namely, the mean of the two principal radii of curvature of the cavity \cite{baduri} integrated over solid angle. 
This oscillatory part is given by,
\be
{\tilde \rho} (\nu) = \sum_{p}A_{p}(\nu,L)\cos(\nu l_p /c+ \mu_p)\nonumber
\ee
where the sum runs over the periodic orbits, $p$, of length $l_p$ of the classical counterpart. 
For electromagnetic waves \cite{bal,blo2} periodic orbits are the  trajectories of the light rays inside the cavity as dictated by geometrical optics. 
The amplitude $A_{p}(\nu,L)$ and the Maslov index $\mu_p$ can be evaluated explicitly from the knowledge of the classical periodic orbits  (see Appendix in Ref. \cite{baduri}).  
In 3d cavities \cite{bal},
\be
 A_{p}(\nu,L) \propto (L/c)^{1+n/2}\nu^{n/2} \nonumber
\ee
where $n$ is 
the number of axes of symmetry of the cavity ($n =3$ for a sphere). 
Polarization effects also modify this amplitude \cite{bal}: To leading order only periodic orbits with an even number of reflections on the boundary of the 
cavity contribute to the amplitude. Moreover the amplitude 
picks up an additional 
factor $2$ ($2cos \alpha$) for
planar (non planar) orbits due to polarization effects. For a given periodic orbit characterized by $m$ reflections, 
\be
2\cos \alpha = {\vec w}\cdot {\vec w_m} +{\vec w'}\cdot {\vec w'_m} \nonumber
\ee
where 
${\vec w},{\vec w'}$ are two orthogonal unit vectors transverse to the direction of propagation of the light ray transforming into  ${\vec w_m},{\vec w'_m}$  after $m$ reflections. The angle $\alpha$ is usually referred to as the polarization angle.  

We note that in this approach $\rho(\nu)$ is effectively written as an expansion in power of the parameter ${\lambda \over L}$ with $\lambda = {c \over \nu}$.  
The limits of applicability of this formalism are thus restricted to a range of sizes $L$ and frequencies $\nu$ such that ${\lambda \over L} \ll1$.

The main goals of this
paper are: a) to provide a detailed account of the impact of finite size effects on the blackbody radiation, b) to determine the 
range of parameters for which such corrections are accessible to experimental verification, c) to propose potential applications of these results.

\section{Corrections to Planck's and Stefan-Boltzmann's laws.}
Finite deviations from the Planck's law Eq.(\ref{a}) are only due to $\rho_{\cal C}$ and ${\tilde \rho}$. As a consequence, the 
modified Planck's law can be written as,
\be
\label{a}
{u(\nu) \over u_V} =1 + a_1\frac{c^2}{L^2\nu^2}+\frac{c^{2-n/2}}{(\nu L)^{2-n/2}}\sum_p {\tilde A}_p \cos(l_p \nu / c+\mu_p).
\ee
where $a_1 \propto {{\cal C} \over L}$ is a dimensionless coefficient that depends on the mean curvature ${{\cal C} \propto L}$ of the cavity, ${\tilde A}_p = A_p c/L$ is the dimensionless amplitude corresponding to a single periodic orbit 
of length $l_p$.  
In the case of symmetric cavities like spheres $(n=3)$ and rectangles $(n=2)$ these coefficients can be obtained easily (see below and \cite{baduri}).  For chaotic cavities ($n =0$) it is much harder to get explicit analytical expressions for $A_p$ except in the case of the shortest periodic orbits \cite{richter}. 

In order to fully determine $u(\nu)$ we have yet to set a cutoff in the sum above. 
From a physical point of view it is evident that periodic orbits longer than 
the length $l_{esc}$ associated to the typical time that a ray stays in the blackbody cavity before escaping through the aperture cannot contribute significantly to $u(\nu)$. 
The explicit expression of the cutoff function depends strongly on the symmetries of the cavity. 
For chaotic cavities ($n=0)$, the contribution to Eq.(\ref{a}) of a light ray of length $l_p$  is weighted by the probability  
that this ray does not escape through the aperture,  
$P_{esc}(l_p) \approx e^{- l_p/l_{esc}}$ with $l_{esc} \sim { {2V\pi} \over {A_{ape}}}\gg l_{min}$, where $V$ is the volume of the blackbody, $l_{min}$ is the length of the shortest periodic orbit and $A_{ape}$ is the surface of the aperture \cite{jala}. 
By contrast for symmetric cavities ($n =2,3$) the cut-off function has power-law tails \cite{jala}.
%The same applies to orbits longer than the coherence length no matter the cause of the breaking of phase coherence \cite{patricio}.      
%We note that in the range of paramters of interest $k L \gg 1$ and consequently fluctuations in integrable 
%grains can be of orders of magnitude larger than in chaotic grains.
% In the range of volumes of interest, namely, $ c/\nu \ll L \ll \infty$,  
%$(L k)^{3/2} \gg 1$. For instance, for $L = 10 \lambda$,  
%$k^{3/2}L^{3/2} \sim 200$. 

The Stefan-Boltzmann's $R(T) = \sigma T^4$ law  is also modified by finite size effects.  
After integrating $u(\nu)$,
\be
R(T) = {\tilde R}(T) +{R_{\cal C}}(T) + \sigma T^4
\ee
where 
\be
{R_{\cal C}} (T) = {1\over {72 \pi}} {{\cal C} \over {L^3 \hbar}} (Tk_B)^2\propto 1/L^2 \nonumber
\ee
 comes from the curvature term $\rho_{\cal C}$ in the spectral density.  
The contribution coming from the fluctuating part of the spectral density  ${\tilde R}(T)$  
can be written exactly in terms of polygamma functions. This expression is rather cumbersome but 
 in the limit $ l_{min} > {{\hbar c} \over {k_B T}}$ (relevant to experiments) simplifies considerably,  
%For chaotic cavities with no symmetries a simple calculation shows,
%\be
%{\tilde R}(T) \approx \frac{\hbar c}{V}\sum_{p,r}\frac{A_{p,r}(L)}{(\tau_p r)^2}\left(1- \left(\frac{\tau_p r/\tau_T}{\sinh(\tau_p r/\tau_T)}\right)^2 \right)
%\ee
%where $\tau_T = 2\hbar/k_BT$ and $r$ counts the number of repetitions of a periodic orbit of period $\tau_p$.
 %For $\tau_T \ll \tau_{min}$ ($\tau_{min}$ the shortest periodic orbit) the second term is exponentially small and ${\tilde R}(T) \approx \frac{\hbar c}{V}\sum_{p,r}\frac{A_{p,r}(L)}%{(\tau_p r)^2}$.  In the range of parameters of experimental interest we shall see $\tau_T \ll \tau_{min}$ is verified
%
\be
\label{r0}
{\tilde R}(T) \approx b_1\frac{c k_B T}{L^3} + b_0 \frac{\hbar c^2}{L^4}
\ee
where the dimensionless coefficients $b_0$ and $b_1$ depend on the number, $n$ of symmetries axes of the 
cavity. The case $n=3$, the sphere, will be discussed in detail later on. 
For $n=0, n=2$,
\be
b_0 =-\frac{n+2}{4}\left(\frac{-2\pi L}{c}\right)^{1+n/2}\sum_{p,r}\frac{A_{p,r}(L)f(\mu_p)}{(\tau_p r)^{2+n/2}} \nonumber
\ee
 with 
$f(\mu_p) =\cos(\mu_p)+(\sin(\mu_p) - \cos(\mu_p))n/2$,  
 \be
b_1 = \left(\frac{2\pi L}{c}\right)^{n/2}\sum_{p,r}\frac{A_{p,r}(L)g(\mu_p)}{(\tau_p r)^{1+n/2}}\nonumber
\ee
 with $g(\mu_p) =-\sin(\mu_p)-(\sin(\mu_p) - \cos(\mu_p))n/2$. Similar expressions are found for $n=1$.
The amplitude $A_{p,r}$ describes the contribution of a single periodic orbit of period $\tau_p = l_p/c$ repeated $r$ times.  
%In the inset of Fig 1. we show explicit results for the case of a sphere $k=3$. 
 We note that: a) this sum is convergent even without including a cutoff function $P_{esc}(l_p)$ due to the aperture, b) the largest contribution to the sum comes from the shortest periodic orbits (light rays), c) the Maslov indexes $\mu_p$ and amplitudes $A_{p,r}$ for the shortest periodic orbits can be computed analytically \cite{baduri}, d) for $T \to 0$ and fixed $L$ 
our semiclassical formalism fails since the maximum of $u(\nu)$ is in the region $\lambda / L > 1$. 
%A similar calculation for cavities with $k$ symmetry axis yields in the $\tau_T \ll \tau_{min}$ limit,

\section{Fluctuations in cavities with no symmetry axes.} 
An explicit analytical determination of ${\tilde R}(T)$ or the oscillating part of $u(\nu)$ involves the knowledge of periodic orbits (light rays) of different lengths. This determination is only straightforward in highly symmetric cavities with several symmetry axis.  
However semiclassical techniques based on that fact that the classical dynamics is ergodic \cite{patricio} permit an estimation of these deviations in the case of chaotic cavities with no symmetries. 
We first study  $\langle {\tilde R^2} \rangle \propto \sum_{p,p',r,r'} A_{p,r}A_{p',r'} $ where the average is over cavities characterized by the same typical length $L$ and with no axes of symmetry.
To leading order, the double sum \cite{baduri} above is given by the diagonal ($p = p'$) term,
\be
\langle {\tilde R^2} \rangle \approx \frac{\hbar^2 c^2}{4 L^6}\int_{\tau_{min}}^\infty d\tau \frac{K(\tau)}{\tau^4}+  \frac{c^2k_B^2T^2}{L^6}\int_{\tau_{min}}^\infty d\tau \frac{K(\tau)}{\tau^2}
\ee
where $K(\tau) = \sum_p A^2_{p,r}\delta(\tau -r\tau_p)$. The lower limit of integration 
corresponds to the period of the shortest periodic orbit  $\tau_{min}= l_{min}/c$. In the 
region around $\tau_{min}$ the function $K(\tau)$  is a collection of well separated peaks located whose exact positions positions depend on the form of the cavity \cite{patricio}. However for longer periods the number of orbits (light rays) with similar periods increases dramatically (in cavities with no symmetries) and  $K(\tau) = 2\tau$ for all chaotic ($n=0$) cavities \cite{patricio}.  
Finally for $\tau > \tau_H$ ($\tau_H = \rho_V(\nu)$ is the Heisenberg time), $K(\tau) = \tau_H$ \cite{patricio}.  Since our purpose is only to estimate the magnitude of the fluctuations we follow \cite{lebflu} and assume that the non universal part of $K(\tau)$ is fully described by the contribution of the 
shortest periodic orbit.  
Within this approximation it is straightforward to show that,
\be 
 \langle {\tilde R^2} \rangle \approx  \frac{\hbar^2 c^4}{4 L^6 l^2_{min}} + \frac{c^2k_B^2T^2}{L^6}\log(\tau_H/ \tau_{min}).
\ee
 We note this result is valid provided that $l_{esc} \gg l_{min}$. Otherwise one would have to multiply the contribution of each periodic orbit by its probability to stay inside the cavity.  

A similar calculation of the fluctuations $\delta u(\nu) $ of the density of energy $u(\nu)$ with respect to $u(_V(\nu)$ in a cavity with no symmetries shows that, 
\be
\frac{\langle \delta u(\nu)\rangle^2}{u^2_V} = \frac{c^2}{\nu^2l^2_{esc}}.\nonumber
\ee
 The dependence of the fluctuations with the cutoff distance $l_{esc}$ is expected since the sum over orbits in $u(\nu)$  Eq. (\ref{a}), unlike $R(T)$ Eq.(\ref{r0}), does not really converge. Without a cutoff $u(\nu)$ is a series of isolated peaks at the natural frequencies of the cavity.    
%We expect a similar result (replacing $l_{esc}$ accordingly) if the cutoff in the sum is induced by other mechanisms such as decoherence or a finite resolution of the apparatus.    

\section{An example: A spherical blackbody.}
The case of a spherical blackbody ($n =3$) is of special interest since it can be solved exactly. In addition finite size corrections are stronger due to the high degree of symmetry of the sphere (the study of a cubic blackbody was carried out in Ref.\cite{case},  the non oscillatory corrections in other simple geometries were investigated in \cite{baltes}).
Small multipolar corrections to the spherical shape \cite{creagh} are described
 with the same periodic orbits that in the spherical case but
adding an additional cutoff in term of Fresnel integral that smoothly modulates the
 amplitude and phase of the fluctuations.
The effect of small, non overlapping bumps    
Ref.\cite{Pav98} 
is only to suppress periodic orbits of the sphere longer than a certain cut-off related to the 
typical size of the bump. 

%The organization of the paper is as follows: First we explicitely compute the blackbody density
% of energy for a spherical cavity. This result will be then compared with a chaotic cavity of similar size.
%Then we discuss different mechanisms leading to the 
% damping of the oscillations caused by the experimental broadening of the linewidth. Finally we discussed 
% application of our results in sonoluminiscence and cosmology.    
\begin{figure}[ht]  
\includegraphics[width=1.0\columnwidth,clip]{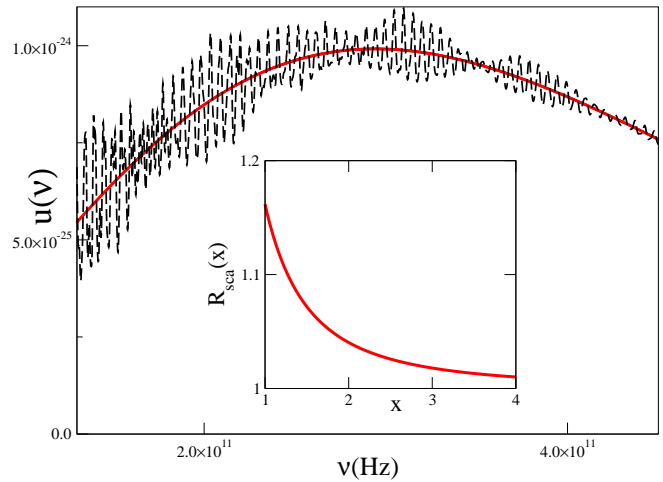}
\caption{Density of energy $u(\nu)$ in $J/m^3$ of a blackbody versus frequency. 
The solid line is the Planck's law.
The dashed line is the analytical prediction for a spherical blackbody Eq.(\ref{a}) and Eq. (\ref{sfe}). 
Finite size effects are clearly visible.
The parameters chosen ($r_0 = 2cm$, $T = 5K$) fall within the region accessible to experiments (see text). In order to reproduce FIRAS resolution only rays shorter than $l_{cut}=2 \times 10^2  c/\nu$  are included in Eq. (\ref{a}).
In the inset we plot $R_{sca}(x) = R(T)/\sigma T^4$ with $x = Tr_0 k_B/\hbar c$.}
\label{callisto}
\end{figure} 
For this geometry the density $\rho(\nu)$ \cite{blo2} is known explicitly.
The closed periodic orbits are given by planar
regular polygons with a even number of vertexes along a plane containing the diameter. 
No polarization angle appears in our expressions due to the 
 planarity of the orbits. The length $L_{p,t}$ of the trajectories is given
by $L_{p,t}=2pr_0\sin(\phi)$ where $r_0$ is the radius of the sphere, $p$ is the number of
vertexes of the polygon and $\phi=\pi t/p$ with $t$ being the number of turns
around the origin of a specific periodic orbit.  With these definitions, 
\begin{eqnarray} 
\label{oscR}
\rho(\nu)  =\rho_V(\nu) + \rho_{\cal C} + {\tilde \rho}(\nu)
\end{eqnarray}
with 
 \be
\label{sfe}
\frac{{\tilde \rho}(\nu)}{V} =-\frac{2\pi}{c}\sum_{t=1}\frac{3\pi \nu}{c r_0t\pi^2}\sin(8\pi\nu tr_0/c)+\\ \nonumber
+\frac{6\pi}{c}\sum_{t,p>2t}(-1)^t{{\left(\frac{2\pi \nu}{c}\right)^{3/2}\sin(2\phi)}}\times \\ \nonumber
\sqrt{\frac{\sin(\phi)}{r_0\pi^3 p}}
\sin((p+3/2)\pi/2+2\pi \nu pr_0\sin(\phi)/c)
\ee
and $\rho_{\cal C} = \frac{8r_0}{3 c}$.
The sum above will be effectively cutoff due to 
the finite aperture that make the cavity open.  We postpone its study to the next section in which we address the experimental detection of these effects. 

Corrections to the Stefan-Boltzmann's law can be evaluated explicitly by integration over $\nu$. 
In the limit ${l_T={{\hbar c} \over {k_B T}}} < l_{min}$, 
\be
\label{r}
R(T) = \sigma T^4 + \frac{1}{12 \pi r_0^2 \hbar }(Tk_B)^2 +  b_1 k_BT c /r_0^3-b_0\hbar c^2/r_0^4
\ee
where  $b_1 \approx 0$ (due to the fact that only orbits with an even number of vertexes contribute to the sum),  $b_0  \approx 
\frac{3}{512\pi^3} \sum_t \frac{1}{t^3}\approx 2\times 10^{-4}$ . 
The second term corresponds to the curvature contribution $\rho_{\cal C}$.  
The last two terms come from the oscillating part of the spectral density ${\tilde \rho}(\nu)$.
 The sum is dominated by the first terms corresponding  to the contribution of the shortest periodic orbits
and consequently it is quite insensitive to the cutoff $l_{esc} \gg l_{min}$. 

In order to examine the importance of these finite size effects we rescale $R(T)$,
 \be
R_{sca}(x) = {R(T) \over {\sigma T^4}} = 1+ {5 \over {\pi^3 x^2}}+{{60 b_0} \over {\pi^2 x^4}}
\ee
 with $x = r_0T k_B /c\hbar$.   This is an expansion in $1/x$ valid for $x \gg 1$. 
The most interesting region (see inset  Fig 1.) to detect finite size corrections is thus $x \gtrsim 1$. We note that since we go up to fourth order the expansion is still accurate for $x \sim 2$.
 Values of $r_0,T$ compatible with this condition are within but close to the limit
 accessible to experiments ($r_0 \geq 0.5 cm$ and $T \geq 2 K$).    
We note that a)  The dependence of $R(T)$ with $T$ and $L$ for a spherical cavity  Eq. (\ref{r}) agrees with the general results obtained earlier Eq.(\ref{r0}). Only the dimensionless coefficients in front of Eq. (\ref{r0}) are cavity dependent,  b) a similar expansion of $R_{sca}$ valid for $x < 1$ can also be carried out. However we could not find a region of parameters ($T,r_0$ with $r_0 > c/\nu$) accessible to experimental verification. 

\section{Experimental detection.}
We determine the range of parameters in which is
feasible to build a spherical blackbody and examine whether finite size corrections in this region could be detected with the 
current technical capabilities. 
%We now describe under what conditions
%finite size corrections to the Planck law could be detected experimentally.

We first ask about the  
 optimal radius for experimental verification. In general, the larger
 the radius the more difficult it is to keep constant and stable
the temperature. This constraint excludes cavities with 
 radius $r_0 \gg 0.2 m$. On the other hand in cavities with $r_0 \ll 1cm$
it is hard to control the geometrical shape. Furthermore, the flux of 
 energy through the aperture is too small to be measured with the accuracy 
 needed.  The optimal radius can be estimated to be $r_0=0.5-7cm$.  We look at wavelengths $\lambda = c/\nu$ 
($r_0\sim 8-30 \lambda$) such that the semiclassical formalism is accurate, namely, ${ \lambda \over {r_0}} \ll 1$. 
Finally the optimal temperature is such that for the range of wavelengths 
 of interest, $\lambda \sim 0.5-2mm$, the density of energy $u(\nu)$ is a maximum.
These conditions are satisfied for $T\sim 2.5-6K$ which is accessible by cooling the cavity with Helium liquid. 
%The reason to use the 
% Wien's law to fix the blackbody temperature is simple:typically wavelength around the 
% maximum of intensity are much easier to de
Thus we propose that the optimal experimental setting is a spherical 
blackbody with $r_0 \sim 0.5-7cm, T \leq 5K$, and  $\nu = c/\lambda \sim 1.5 - 6 \times 10^{11} s^{-1}$.
In addition in order to observe the effect of the periodic orbits ($l_{esc} \gg l_{min}$) the area of the aperture must verify $A_{ape} \ll {2 \over 3}\pi^2r_0^2$. 

In order to proceed we must choose an apparatus of measurement.
%We want to show that the detection of finite size corrections is 
% already possible with current state of the art technology.   
For our purpose it is of paramount importance that the radiation of the blackbody at $T\sim 5K$ can be measured with the highest precision. We also request
that the resulting flux of energy Eq.(\ref{a}) through the aperture can be measured with a precision much larger than the strength of the predicted finite size 
 corrections. Finally we require the apparatus to be precise enough to discern comparatively close 
wavelength. In order to observe fluctuations in $u(\nu)$  it is required that the apparatus can discern frequencies  $\Delta(\lambda)$ comparable with the contribution to $u(\nu)$ of the shortest periodic orbit.  In our case this corresponds to  $\Delta(\lambda)/\lambda \leq 0.05$. 

After a careful research, we conclude that instruments designed to measure the cosmic microwave background are the best suited 
 for our purpose. We focus our analysis on FIRAS \cite{firas}
 apparatus on board of COBE satellite though other differential devices to measure the cosmic microwave background could be used as well. 
By differential it is meant that an internal calibrator  
 nulls the external signal coming either from the sky or from the external calibrator. This 
 mechanism is responsible for the FIRAS ability to detect small deviations from a blackbody source
(ICAL in FIRAS).  
The FIRAS spectrometer is designed to measure deviations from  
a blackbody spectrum in the region $\lambda=0.1-10mm$. The intrinsic
frequency resolution is $\Delta(\nu)/\nu \sim 5 \times 10^{-3}$. This implies that the contribution of periodic orbits (light rays)  $l_p \gg l_{cut} = 2\times 10^{2}c/\nu$ will not be detected by the apparatus.   
In the range of frequencies of interest the equivalent noise power $\sim 4\times 10^{-15}W/\sqrt{Hz}$ is less than $1 \%$ of the measurement. The blackbody temperature can be set in the range $T = 2-10K$ with a precision of a few $mK$.  
%For a 
% throughout review about FIRAS we refer to \cite{firas}
In Fig.1 we plot $u(\nu)$ for a spherical cavity for a set of parameters ($r_0 = 2cm$, $T=5K$ and $l_{cut} = 2\times 10^{2}c/\nu$) accessible to experimental verification.
From the figure it is clear that an apparatus with FIRAS specifications 
is capable to detect finite size corrections in spherical blackbody cavities. Similar results will be obtained for other symmetric cavities such as cylinders or cubes. 
% For the sake of comparison we have also plot
% the prediction for a chaotic cavity of the same size. In this case 
%the deviation from Planck's law is less than $1 \%$.
%These fluctuations cannot however be observed in  
%the effective cavity formed by the sky horn and the external FIRAS calibrator 
%since the latter is made of a highly absorbing material that suppresses even the contribution of the shortest periodic orbits (light rays).

We note that a condition for the experimental observation of these effects is that 
the radiation is coupled out of the cavity in such a way the original modes of the cavity are not seriously affected.  
Previously we have shown that the aperture of the blackbody acts just as a natural cutoff for long periodic orbits. Therefore this is not a problem for the experimental 
verification of our results. Indeed in recent experiments
\cite{richter} it has been possible to measure the natural modes of a cavity with a microwave source with a precision enough to even test quantitatively semiclassical estimation for the number of modes.

%  apparatus but replacing the 
%sky horn by a spherical black body cavity with appropriate (see above) radius and temperature.  
\section{ Applications: Sonoluminescence and the cosmic microwave background.}
 Differential apparatus as FIRAS compare
the sky signal with some calibrator on board which is considered a 'perfect' blackbody.
These calibrators are effective blackbody cavities which may be affected 
 by the finite volume corrections reported in this letter. These corrections may 
 very well be confused with those coming from the sky.
 Thus on board calibrators (or cavities used as reference sources in the measurement, for instance, of the spectral emissivity of solid samples \cite{baltes}) should be carefully designed to minimize these effects.

We speculate that finite size corrections may also play a role in cosmological problems. The CMB
 observed today had its origin in photons from the 
last scattering surface. For angular separations larger than 
$1.5$ degree, temperature fluctuations in the CMB comes from regions that 
have never been in thermal contact. The origin of these fluctuations is supposed
 to date back to the inflation time.
 We speculate that finite size effects may play some role in the generation of these fluctuations.
 In order to determine the true 
relevance of this effect it would be necessary to: a)understand 
the physical mechanisms such as decoherence that may lead to a suppression of these effects, b) study these deviations in a non Euclidean cavity resembling the 
global geometry of the universe. It is encouraging that in Euclidean cavities it can be shown that the shape of the 
finite size corrections to the blackbody's laws are not modified as universe expands.

Sonoluminescence, the transduction of sound into light, \cite{weninger,revson},
occurs when the pulsations of an almost spherical bubble of gas in water (or other substances \cite{camara}) produced by a standing sound wave  
attains sufficient amplitude so as to emit periodic picosecond flashes of light.  In a certain range of frequencies 
the spectrum is consistent with a blackbody \cite{camara, weninger} of typical size $r_0 \sim 1\mu m$ and temperatures $T \sim 10^4 K$  (for bubbles of Xenon in sulphuric acid $r_0 \sim 3.9\mu m$ and $T \sim 7000K$).  In this range of parameters finite size effects should be observable.  However we note in this case the blackbody is not the perfecting conducting cavity with an aperture studied previously but rather a material of refraction index $n_1$ immersed in a medium with refraction index $n_2$.  A natural question to ask is whether the 
analytical results of the previous section are also applicable in this situation. The answer is affirmative. The only difference is the way in which the cutoff of long periodic orbits is defined. 
For $n_1 > n_2$ the blackbody is effectively an open cavity.  A given periodic orbit (light rays) inside the material of refraction index $n_1$ will contribute to the spectral density provided 
that the angle of incidence $\theta > \theta_c$ (measured
from the normal) with $\theta_c = \arcsin (n_2/n_1)$.  Periodic orbits such that $\theta < \theta_c$ do not contribute since the light ray escapes to the medium with $n_2$ \cite{bogodi}. 
%In addition it is unclear to us whether the analytical results for the density of energy in the cavity like blackbody of the previous section hold unmodified in this case.  Obviously the %smooth part of the corrections, the curvature term,
 %should be the same but the oscillatory part could be different. 
%Despite 
 %these difficulties we consider sonoluminescence a very promising area to 
%observe finite size effects in the blackbody radiation.  

In conclusion, by using semiclassical techniques, we have derived explicit expressions for the finite size corrections
of the blackbody radiation's laws as a function of the temperature, size and shape of the cavity and area of the aperture. 
We have also shown that the experimental detection of these finite size corrections is within the 
 reach of current experimental capabilities and may be of relevance in sonoluminescence.

\acknowledgements
I thank Patricio Leboeuf for suggesting me to look at this problem. I appreciate conversations with Keith Weninger,  Jim Peebles and Juan Diego Urbina. I 
warmly thank referees for insightful comments and suggestions.
I acknowledges financial support from a Marie Curie Outgoing
Action, contract MOIF-CT-2005-007300.


\begin{thebibliography}{9}
\bibitem{bal}R.Balian and B. Duplantier, Ann. Phys. {\bf 104}, 300 (1977).
\bibitem{blo1} R. Balian and C. Bloch, Ann. Phys. {\bf 64}, 271 (1971).
\bibitem{blo} R. Balian and C. Bloch, Ann. Phys. {\bf 60}, 401 (1970).
\bibitem{blo2} R. Balian and C. Bloch, Ann. Phys. {\bf 69}, 76 (1971).
\bibitem{baduri}M. Brack, R.K. Bhaduri, {\em Semiclassical Physics}, Addison-Wesley, New York, (1997).
\bibitem{patricio} P. Leboeuf, Lect. Notes Phys. {\bf 652}, Springer, Berlin Heidelberg 2005, p.245, J. M. Arias and M. Lozano (Eds.).
\bibitem{richter}	C. Dembowski, et al.,
Phys. Rev. Lett. {\bf 89}, 064101 (2002).
\bibitem{jala}R.A. Jalabert, cond-mat/9912038.
 \bibitem{lebflu} H. Olofsson,et al., arXiv:0704.2310;O. Bohigas and P. Leboeuf, Phys. Rev. Lett. {\bf 88} (2002) 092502.
%\bibitem{ric}C. Dembowski et.al, Phys. Rev. Lett. {\bf 89} 064101 (2002).
\bibitem{case}K.M. Case and S.C. Chiu, Phys. Rev. A {\bf 1} 1170 (1970);H.-T. Elze and W. Greiner, Phys. Lett. {\bf B179} (1986) 385. 
\bibitem{baltes}H.P. Baltes and F.K. Kneubuhl, Phys. Lett{\bf 30A}, 360 (1969);H.P. Baltes and F. K. Kneubuhl,
Hel. Phys. Acta {\bf 45}, 481 1972.
\bibitem{creagh}P. Meier,et al.,  Z. Phys. D {\bf 41} 281 (1997).
\bibitem{firas} J.C. Mather,et al., Ap.J {\bf 512} 511 (1999).
%\bibitem{weninger1}G. Vazquez, C. Camara, S. Putterman and K. Weninger, Opt. Lett. {\bf 26} 575 (2001).
\bibitem{camara}S. D. Hopkins et al., Phys. Rev. Lett. 95, 254301 (2005).
\bibitem{weninger}G. Vazquez,et al., Phys. Rev. Lett. {\bf 88} 197402 (2002);G. Vazquez, C. Camara, S. Putterman and K. Weninger, Opt. Lett. {\bf 26} 575 (2001).
\bibitem{revson}B.P. Barber, et al., Phys. Rep. {\bf 281} 66 (1997).
\bibitem{bogodi}M. Lebental, et al., Phys. Rev. A 76, 023830 (2007). 
\bibitem{Pav98} N. Pavloff and C. Schmit, Phys. Rev. {\bf B 58}, 4942 (1998).
%\bibitem{ric1}H.Alt et. al., Phys. Rev. Lett. {\bf 79} 1026 (1997).
%\bibitem{koch}L.Sirko and P.M.KochPhys. Rev. E 54, R21-R24 (1996).  
\end{thebibliography}
\end{document}